\documentclass[]{aa}
\usepackage{natbib}
\usepackage{psfig}
\def\EBV{\mbox{E$_{\rm B-V}$}}

\def\HH{\mbox{H$_2$}}
         
\def\nH2{{\rm n}({\rm H}_2)}
\def\NH2{{\rm N}({\rm H}_2)}
\def\pccc{~{\rm cm}^{-3}} \def\pcc{~{\rm cm}^{-2}}

\def\Tstar#1 {\mbox{${\rm T}_{\rm #1}^*$}}
\def\Tsub#1 {\mbox{$T_{\rm #1}$}}
\def\TK  {\Tsub K }

\def\Tsp {\Tsub sp }

\def\Texc {\Tsub exc }

\def\Tcmb{\Tsub cmb }

\def\arcsec{\mbox{$^{\prime\prime}$}} 
\def\arcmin{\mbox{$^{\prime}$}}

\def\p{\mbox{$^+$}}

\def\hcop{\mbox{{HCO\p}}}

\def\cch{\mbox{C$_2$H}}

\def\h13cop{\mbox{{H$^{13}$CO\p}}}

\def\c3h2{\mbox{C$_3$H$_2$}}
 
 \def\R0{R$_0$}

\def\ddeg{{}^\circ\kern-.1em}

\def\kms{\mbox{km\,s$^{-1}$}}

\def\E#1 {$10^{#1}$}
\def\E#1 {E{#1}}
\def\P#1,{$\nH2\TK~=~#1\times~10^4\pccc$~K}
\def\ec#1,#2,#3,{#1\,(#2)\E{#3}}

\def\zoph{$\zeta$ Oph}

\def\H3{\mbox{H$_3$}}

\title{Mm-wave \hcop, HCN and CO absorption toward NGC1052}
\author{H. Liszt\inst{1}\ and R. Lucas\inst{2} }
\institute{National Radio Astronomy Observatory,
           520 Edgemont Road,
           Charlottesville, VA,
           USA 22903-2475
\and       Institut de Radioastronomie Millim\'etrique,
           300 Rue de la Piscine,
           F-38406 Saint Martin d'H\`eres,
           France}

\begin{document}
\date{received \today}
\offprints{H. S. Liszt}
\mail{hliszt@nrao.edu}
\abstract{
We used the Plateau de Bure Interferometer to observe $\lambda$3mm 
J=1-0 absorption lines of \hcop , HCN and CO toward the core of
the nearby elliptical, megamaser-host galaxy NGC1052.  The lines
are relatively weak, with peak optical depths 0.03 for \hcop\ and 
HCN and 0.1 for CO.  Nonetheless the inferred column density of 
molecular gas 2N(\HH) $\simeq 5\times10^{21}~\pcc$ is consistent 
with the degree of reddening inferred toward the nucleus from 
observations of the Balmer series of hydrogen.  Mm-wave absorption 
line profiles are somewhat broader than those of H I and OH, perhaps 
because lower free-free opacity at mm-wavelengths exposes higher-velocity 
material nearer the nucleus.  Overall, the OH/\hcop\ ratio in NGC1052 is
as expected from the strong relationship established in local diffuse 
clouds but the optical depth ratio varies strongly over the line 
profiles.  Similar variations are also seen toward Cen A, which has 
very different line ratios among H I, OH and \hcop\ for very nearly 
the same amount of OH absorption.  
}
\titlerunning{Molecuar gas in NGC1052}
\maketitle

\section {Introduction.}

\begin{figure}
\psfig{figure=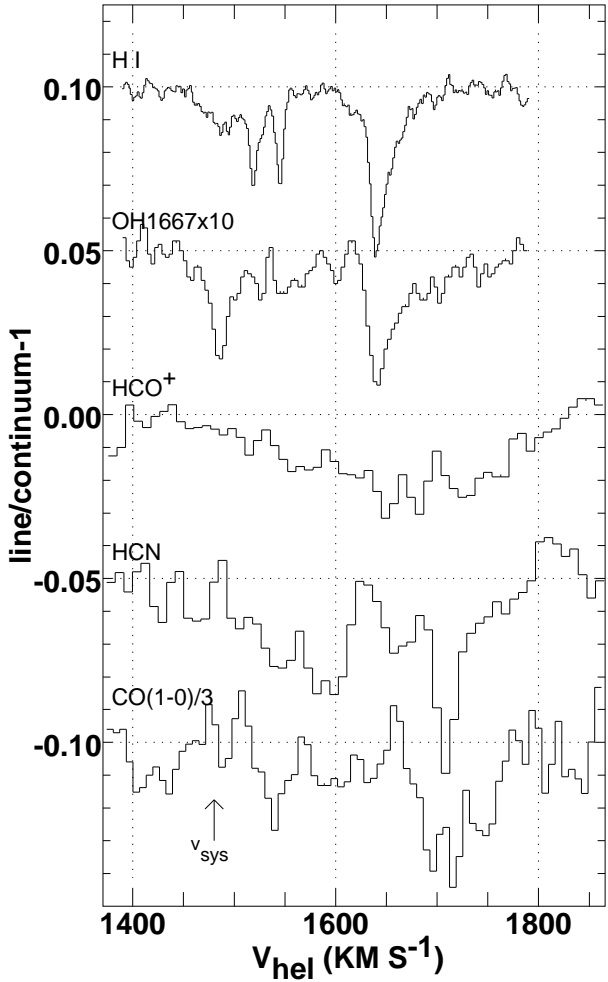,height=13cm}
\caption[]{Absorption profiles toward NGC1052.
The data are presented as scaled, offset, line/continuum
ratios.  The H I and OH profiles at top are 
the "total" VLB profiles from \cite{VerRos+03}
representing spatially-resolved data averaged over
some 15-20 mas or 1.5 - 2.0 pc at one epoch.  Profiles of the 
J=1-0 transitions of HCN, \hcop\ and CO from the Plateau de Bure 
represent data taken at multiple epochs against the vastly-unresolved
mm-wave continuum, whose spatial distribution is compact and probably
similar to that at lower frequencies, but with a larger contribution 
from regions at smaller radii (see Sect. 3.1 of the text)}
 
\end{figure}

NGC 1052 is a nearby elliptical galaxy (v$_{sys}$ = 1480 \kms; 
1\arcsec = 102.5 pc for H$_0$ = 70 \kms\ Mpc$^{-1}$) with a 
variety of distinguishing nuclear characteristics.  It hosts a 
weak LINER active galactic nucleus (AGN) \citep{GabBru+00} with 
a pc-scale VLBI jet-counterjet structure visible up to 43 GHz 
\citep{VerRos+03}. 
 An \HH O megamaser is seen against the counter-jet 
\citep{ClaDia+98} and NGC 1052 is the only case in which an elliptical 
galaxy is known to host such phenomena: the other elliptical cited by 
\cite{ClaDia+98}, TXSS 2226-184 \citep{KoeHen+95}, has recently been 
shown likely to be a disk+bulge system \citep{FalWil+00}.  
On kpc-scales one finds a dust lane and 
a pair of complementary counter-rotating nuclear gas disks 
\citep{PlaBou96} as well as a double radio 
source \citep{Wro84}.  An extended H I envelope about
NGC 1052 was found by \cite{vanKna+86} and \cite{OmaAna+02}
recently discovered OH absorption against the nucleus as well.  
 OH absorption over a wider range of velocity
 was found by \cite{VerRos+03}, who also present a comprehensive 
study of H I absorption. 

Weak absorption against the nuclear continuum complicates 
interpretation of the H I emission over the optical disk of the 
galaxy but the column densities seen in 21cm emission at relatively
low resolution 
( 1\arcmin $\approx 6$ kpc)
 are moderate, of order N(H) $\simeq 10^{20}~\pcc$ 
(Fig. 5 of \cite{vanKna+86}).  The hydrogen column densities inferred 
from the presence of OH are larger
than those derived from the H I emission, consistent with obvious 
optical obscuration and the inference 
of nuclear reddening (albeit with substantial uncertainty) 
$\EBV=0.42\pm0.2$ mag. by \cite{GabBru+00}, implying 
N(H) $\simeq 2.5\times10^{21}~\pcc$ for conditions like those
seen locally in the disk of the Milky Way.

In the Milky Way, the detection of cm-wave OH absorption is a sure 
indicator of the presence of other trace molecules, especially
\hcop\ \citep{LucLis96,LisLuc96}, CO \citep{LisLuc98}. 
\cch\ \citep{LucLis00a}, and (somewhat more occasionally) CN and 
HCN \citep{LisLuc01}, even in diffuse clouds.   Given 
this, we sought and found mm-wave absorption from \hcop, HCN and 
CO toward the nucleus of NGC 1052.  The implications of these
observations, and a comparison with Cen A ( NGC\\,5128;
the only other 
elliptical where the same absorption lines have been observed)
are the subject of this work.  Section 2 describes the 
observations and a discussion is given in Sect. 3.

\section{Mm-wave observations of \hcop, HCN and CO}

The J=1-0 transitions of HCN (88.631 GHz), \hcop\ (89.188 GHz) and CO 
(115.271 GHz) were observed at the Plateau de Bure Interferometer at 
various times during the year 2001. The new spectra discussed here,
shown in Fig. 1, were taken with 2.5 MHz wide correlator channels 
corresponding to 8.46, 8.50 and 6.5 \kms\
for HCN, \hcop\ and CO respectively.  The passbands were much wider 
than the excerpts shown here.  Reduction of such data, where
a strong unresolved background source is also the calibrator, is 
basically a matter of accumulating line/continuum ratios.  
The continuum level, approximately 
0.6 Jy, was not precisely determined, however.  Spectral baselines 
in the data were determined from the feature-free regions alongside 
the spectra.

The column density in the lowest (J=0) level of a simple linear
molecule is related to the integrated optical depth of the J=1-0
transition as

$$ N_0 = {{8.0\times10^{12} \pcc \int\tau_{10} dv}\over
{\mu^2 (1-\exp{(-h\nu_{10}/k\Texc)})}} 
\eqno(1)$$

where $\mu$ is the permanent dipole moment of the molecule, 
$\Texc$ is the excitation temperature of the J=1-0 transition,
$\nu_{10}$ is its frequency and the profile integral is in
units of \kms.  The dipole moments of the species observed here 
are 2.98 (HCN), 4.07 (\hcop) and 0.112 Debye.  Except in fairly 
dense gas, the excitation of HCN and \hcop\ is seen to be very weak 
\citep{LucLis96,LisLuc96} and the 
column density can be reliably determined from observation of 
the J=1-0 line.  Denoting the J=1-0 optical depth profile integral 
as I$_{10}$ (units of \kms ) and taking the excitation temperature 
as 2.73 K,
one finds
N(HCN) = $1.92\times10^{12}~\pcc~$I$_{10}$,
N(\hcop) = $1.02\times10^{12}~\pcc~$I$_{10}$ and
N(CO) = $1.03\times10^{15}~\pcc~$I$_{10}$.  
For an excitation temperature of 10 K,
N(HCN) = $1.31\times10^{13}~\pcc~$I$_{10}$,
N(\hcop) = $6.96\times10^{12}~\pcc~$I$_{10}$ and
N(CO) = $5.96\times10^{15}~\pcc~$I$_{10}$. 

Below we also consider the column density of OH determined
from observation of the 1667 MHz line,  which for LTE conditions
is written N(OH) = $2.24\times10^{14}~\pcc~\Texc~\int\tau dv$.
For diffuse clouds locally this is believed to be a good 
approximation, and the excitation is found to be quite
weak, \Texc-\Tcmb $\simeq 1$ K \citep{LisLuc96,FelRou96}.

\subsection{Dense or diffuse/translucent gas?}

 The present work discusses the observations in terms 
of diffuse/translucent gas of low true optical depth and a 
fully-covered continuum, which seems reasonable given the small 
size of the latter -- at most 2 pc and perhaps much less if the 
core is exposed at/above 90 GHz; see Sect 3.  This provides a 
consistent and rather compact interpretation with a minimum of 
argument and uncertainty.  
However, nothing in the presently-available data 
absolutely refutes the alternative that the gas is dense and has 
higher optical depth and low covering factor, at least in part.  
For instance, the excitation temperature of the CO 1-0 transition 
overall could be 100 K, rather than 6 K, if the expected amount 
of gas phase carbon is to be put in dense, rather than diffuse, 
gas (see Sect. 3). 
Observations of higher-lying mm-wave transitions might help to
settle this issue, though the rapid evolution of the continuum 
with epoch and observing frequency will complicate matters.

\subsection{Borrowed and previously-published spectra}

We are indebted to several authors for providing copies of their
previously-published data.  For reference, they are: toward NGC1052, 
H I and OH absorption profiles from \cite{VerRos+03}, and 
toward Cen A, H I from \cite{SarTro+02}, OH from \cite{vanvan+95} 
and \hcop\ from \cite{WikCom97}.  A datacube of OH absorption 
profiles seen in the Milky Way near Sgr A$^\ast$ originally taken 
by \cite{KilLo+92} is referenced in Sect. 3.  The CO J=3-2 emission 
profile toward Cen A in Fig. 2 is from \cite{Lis01a} and was taken 
at the JCMT with $\simeq$ 14\arcsec\ resolution.

\section{MM-wave absorption toward NGC 1052}

The new mm-wave absorption data are summarized in Table 1 and 
shown in Figure 1.  Several strong and distinct features seen
in OH and H I are not apparent at mm-wavelengths.  The CO
and HCN profiles seem to resemble each other more closely
than the \hcop.  The latter, which has by far the lowest
noise in our dataset, is somewhat broader than the OH, which 
is in turn noticeably more extensive than H I.  As discussed
in Sect. 3.2 (see Fig. 2), the various species seen toward Cen A
exhibit quite similar behaviour.  

 The differences between H I and OH which occur in the two 
ellipticals are quite typical of what is seen locally along 
rather random directions toward extragalactic point radio continuum 
sources \citep{LisLuc96}, reflecting the localization and/or 
segregation of cool and diffuse molecular gas in the nearby ISM.
As such, they do not necessarily signal differences between
nuclear and off-nuclear or dense and diffuse gas toward NGC1052.
Similarly, sharp differences between the absorption profiles of 
\hcop\ and species like HCN \citep{LisLuc01} or CO \citep{LisLuc98} 
arise along local lines of sight because of differing stages of 
chemical maturity in the various diffuse clouds.  However, differences 
between \hcop\ and OH absorption profiles generally do not occur in 
the local ISM and this issue is addressed separately in Sect. 3.1 and 3.3.

The \hcop\ profile integral converted to column density
in the low-excitation limit yields N(\hcop) 
$= 5.5 \times10^{12}~\pcc$, implying 2N(\HH) 
$= 3.6-5.6 \times 10^{21}~\pcc$  for \hcop/\HH\ ratios 
X(\hcop) $= 2-3 \times 10^{-9}$ typical of galactic diffuse 
and translucent gas.   Locally, such a gas column would imply 
a reddening of \EBV\ = 0.6 - 1.0 mag which is consistent
with the inference of \cite{GabBru+00} that \EBV = $0.42\pm0.2$ mag.
The same exercise may be repeated with OH, which in
diffuse and translucent gas has a relatively stable
relative abundance X(OH) $= 10^{-7}$ \citep{LisLuc96,LisLuc02}.  
For an excitation temperature of 4 K, 
N(OH) $= 3.27 \times 10^{14}~\pcc$ implying 
2N(\HH) $= 6.5 \times 10^{21}~\pcc$.  The consistency
of these estimates implies that the  mean OH/\hcop\ 
 opacity ratio is similar to that seen in the Milky Way 
as well (see Sect. 3.2).

\begin{table}
\caption[]{Integrated optical depth: NGC 1052}
{
\begin{tabular}{lcccc}
\hline
Line& $\sigma_{l/c}^1$& low v$^2$ & high v & total \\
\hline
&&\kms &\kms & \kms \\
\hline
H I$^3$ &  0.0017 & 1.34(0.02) & 1.63 (0.02) & 2.97 (0.03)  \\
OH$^3$  &  0.0036 & 0.153(0.008) & 0.213(0.009) & 0.365(0.012) \\
HCN & 0.0086  & 2.89(0.40) & 3.25 (0.40) & 6.15 (0.58) \\
\hcop & 0.0037 & 1.50(1.12) & 4.11 (0.12) & 5.47 (0.18) \\
CO &  0.027 & 3.64 (1.13) & 9.92(1.15) & 13.56 (1.61) \\
\hline
\end{tabular}}
\\
$^1$ $\sigma_{l/c}^1$ is the rms noise in the line/continuum ratio \\
$^2$ v $<$ 1600 \kms \\
$^3$ data from \cite{VerRos+03}
\end{table}

\begin{table}
\caption[]{Integrated optical depth:  Cen A (NGC\\,5128)}
{
\begin{tabular}{lcccc}
\hline
Line&$\sigma_{l/c}$& low v$^1$ & high v & total \\
\hline
&&\kms &\kms & \kms \\
\hline
H I &  0.0022 & 14.03(0.02) & 13.53 (0.02) & 27.43 (0.03)  \\
OH  &  0.0020 & 0.218(0.008) & 0.225(0.013) & 0.463(0.015) \\
\hcop & 0.017 & 15.83(0.18) & 6.34 (0.10) & 22.16 (0.21) \\
\hline
\end{tabular}}
\\
$^1$ v $<$ 565 \kms\ \\

\end{table}

The column density ratio N(HCN)/N(\hcop) = 2.1 is entirely typical of 
local gas \citep{LisLuc01}; this ratio is relatively unaffected by 
assumptions about the overall excitation, although \hcop\ is somewhat
more easily excited in diffuse gas.  By contrast, carbon monoxide is 
unlikely to remain as cold rotationally as the \hcop.  The  CO column 
density is  $1.4 <$ N(CO) $ <  8.1\times 10^{16}~\pcc$
for 2.73 K $<$ \Texc $<$ 10 K.  CO/\hcop\ ratios as large
as that implied by the upper limit CO column density 
are rather high for diffuse gas \citep{LisLuc98}.  Even so,
the CO/H ratio N(CO)/2N(\HH) 
$\la 8.1\times 10^{16}~\pcc/ 5\times 10^{21}~\pcc \simeq 10^{-5} $
is still about one order of magnitude too small to account for
the bulk of the free gas-phase carbon expected for material having
Solar metallicity and carbon depletion onto grains like that 
seen locally, where [C]/[H] = $1.4 \times 10^{-4}$ \citep{SavSem96}.

The absence of most of the expected carbon from CO suggests either 
low metallicity or diffuse molecular gas, see \cite{LisLuc98} or 
Fig. 9 of \cite{LisLuc00}.  However, assuming a gas of lower metallicity 
would probably still not account for all the gas phase carbon, 
because the OH/\HH\ and \hcop/\HH\ ratios used to derive N(\HH) would
presumably be smaller in a gas of lower metallicity, thereby increasing the 
inferred N(\HH).  Nonetheless, there is a somewhat indirect suggestion 
of sub-solar metallicity, found in another context.  \cite{GabBru+00}
noted (see their Sect. 5) that some anomalies in their photoionization
models would be alleviated if the 220 nm bump were not present in the 
$uv$ extinction curve.  This is typical of SMC-like extinction curves 
\citep{PeiFal+91}.

\begin{figure}
\psfig{figure=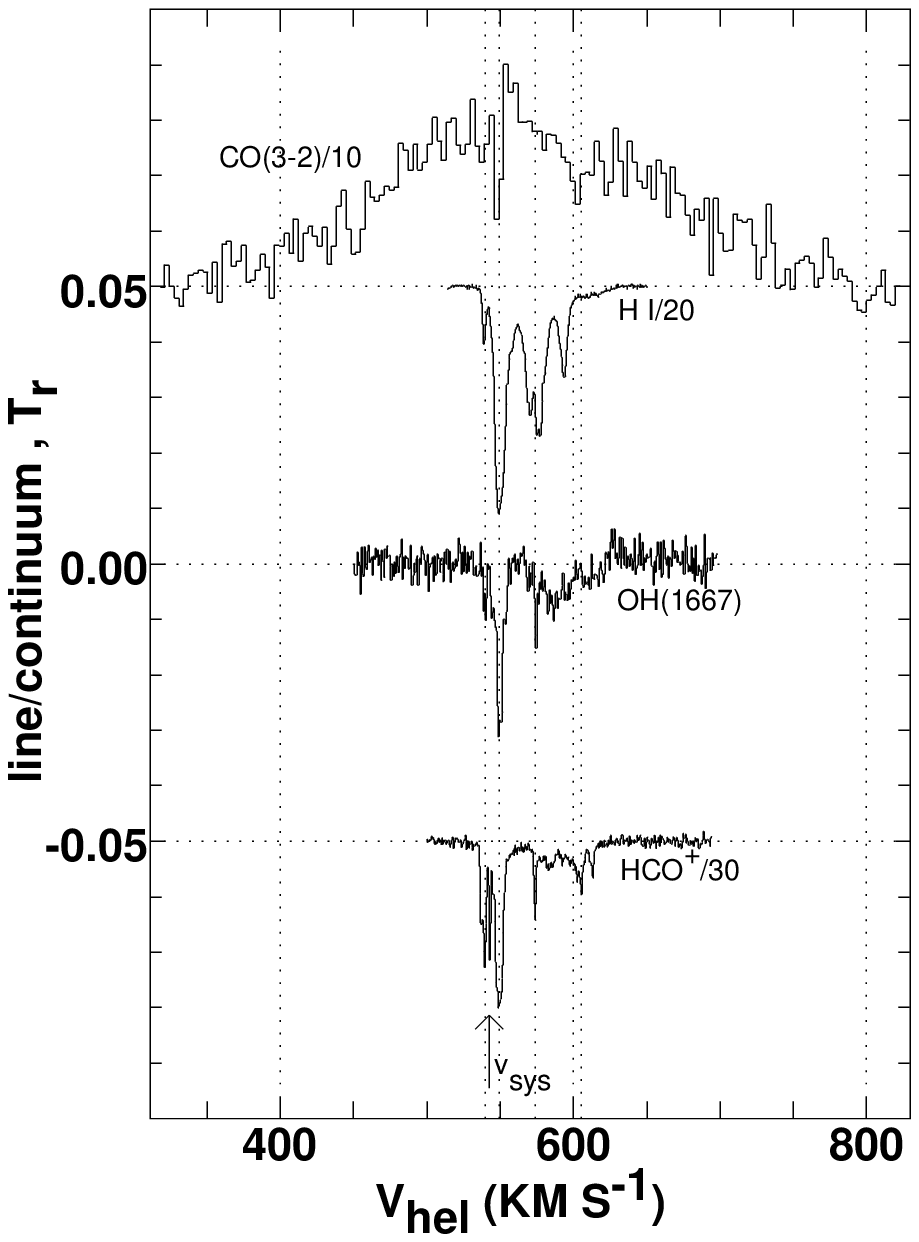,height=11.5cm}
\caption[]{CO J=3-2 emission and various absorption profiles
toward Centaurus A. The CO is the centermost profile in
the study of \cite{Lis01a}. The H I is from \cite{SarTro+02},
OH from \cite{vanvan+95} and \hcop\ from \cite{WikCom97}.
}
\end{figure}

The observed CO J=1-0 optical depth, if it arises in a gas with 
\Texc\ = 10 K, would produce an emission line with a brightness 
temperature 0.5 K and an integrated profile brightness 40 K \kms.  
Combined with a galactic CO/\HH\ conversion factor, such an integrated 
brightness would imply 2 N(\HH) $ \simeq 40 \times 4\times 10^{20}~\pcc$.
This is a factor 2-3 above the estimates of 2N(\HH) from \hcop\ and 
OH.  However, the brightness of any CO J=1-0 emission from NGC 1052 is 
not as large as 0.02 K \citep{WikCom+95} over a 25\arcsec\
beam.  

From these arguments we draw several conclusions.  The excitation
temperature of the CO is probably smaller than 10 K; taking 
\Texc = 6 K, implying N(CO) $= 3.5 \times 10^{16}~\pcc$  would 
lower the integrated brightness by a factor two
 with respect to the estimate in the preceding paragraph,
reconciling the 
estimates of N(\HH) from the various trace molecules.  Secondly, the 
CO gas producing the observed absorption cannot fill more than a 
small fraction of the  10\arcsec - 60\arcsec singledish
telescope beams which have been used to search 
for CO emission; crudely, at most of order 0.02 K/0.25 K.  However, 
at this level, 
the line emission flux into a singledish beam may not be large 
compared to the continuum flux which is being absorbed out by 
the gas we have detected.  That is, 10\% absorption of an 0.6 
Jy continuum, observed by a telescope with 6Jy/K gain, produces 
a negative signal of 0.1*0.6/6 = 0.01 K.  Thus, searches for CO 
emission have probably been compromised somewhat by the same confusion 
between emission and absorption which afflicted the H I synthesis 
of \cite{vanKna+86}.  Off-nuclear searches for CO emission in the
optical dust lane would avoid this problem.

The inferred molecular column densities are much higher than those
of H I seen in emission about the optical disk of NGC 1052, 
N(H I) $\simeq 10^{20}~\pcc$ \citep{vanKna+86}.  However, the H I 
absorption column density implied by the integrated optical depth, 
N(H I) = $5.3 \times 10^{20}$ \Tsp/100 K, is also quite likely to 
be much larger than that seen in the low-resolution emission data.  
In the absence of much higher resolution H I emission 
data, very little can be said about the actual spin temperature 
in the atomic gas, except to note that the amount of H I
absorption which is seen is substantially higher than would 
be produced by extra-nuclear foreground gas having N(H I) 
$= 10^{20}~\pcc$ at a typical \Tsp = 100 K, implying that
the H I absorption does in fact arise nearer the nucleus. 
Conversely, the H I absorption column density is still 
 too small to account for the observed reddening unless 
the spin temperature is extraordinarily high.   It seems more 
likely that the gas associated with the bulk of the occulting 
column density is that sampled by the trace molecules.

\subsection{Kinematics}

If the \hcop\ profile is wider and generally different in shape 
than the OH, such differences are not expected to arise 
chemically -- as when comparing OH or \hcop\ with H I -- but may occur
for any of several other reasons.  The optical depth of the
1667 MHz OH line is subject to excitation effects which produce
rather spectactular results at higher OH column densities 
between $10^{14}- 10^{15}~\pcc$ \citep{vanvan+95} but OH exists
in phases of lower and higher excitation even in diffuse gas
\citep{LisLuc96}.  The OH-\hcop\ comparison is of particular
importance and is discussed separately in Sect. 3.3.

Another effect which may be important in NGC 1052 is the
frequency dependence of the free-free opacity about the
nucleus.  \cite{KamSaw+01} have modelled an asymmetric electron
distribution having a peak free-free opacity of 300 toward 
the core at 15.4 GHz.  Scaling as $\nu^{-2.1}$, the peak opacity 
at 89-115 GHz would be only 4-7.  Thus the innermost parts of 
the jets could become visible in mm-wave absorption (there is a 
pronounced gap in VLB maps up to 43 GHz, see \cite{VerRos+03}), 
perhaps exposing higher-velocity foreground material to 
illumination by background continuum.  

Another possibility for creating differences is the secular
evolution of the blobby knot structure within the VLB jet; the
few years which elapsed between the observations of \cite{VerRos+03}
and our own (from 1997 and 1998 to 2001) most likely witnessed 
substantial evolution in the jet structure.  Spectral index
variations over the jet structure will also result in different
weighting of absorption from foreground gas seen at widely
differing frequencies.

The kinematics of the inner regions are rather poorly understood.  
Redshifted gas appears on both sides of the nucleus, to the West 
in the \HH O maser spots seen against the counterjet and to the 
East and West in H I absorption \citep{ClaDia+98,VerRos+03}.  The 
H$\alpha$ line \citep{PlaBou96}  shows
complementary prograde and retrograde rotating disks on scales
of 1-2 kpc (10\arcsec-20\arcsec) but the velocities of neither
structure (relative to the systemic) approach those seen
in the masers or radiofrequency absorption lines.  This is
analogous to Cen A, where the nuclear disk seen in CO 
emission \citep{Lis01a} has no presence whatsoever in H$\alpha$ 
\citep{BlaTay+87}.  The differences between CO and H$\alpha$ in
the nucleus of Cen A stand in sharp contrast to what is seen
over the larger dust lane, where they behave essentially
identically within the various limitations
of the observational data.  However, the circumnuclear disk in Cen
A is well-described as purely rotating \citep{Lis01a}, which cannot
account for the redshifted absorption features.

\begin{figure}
\psfig{figure=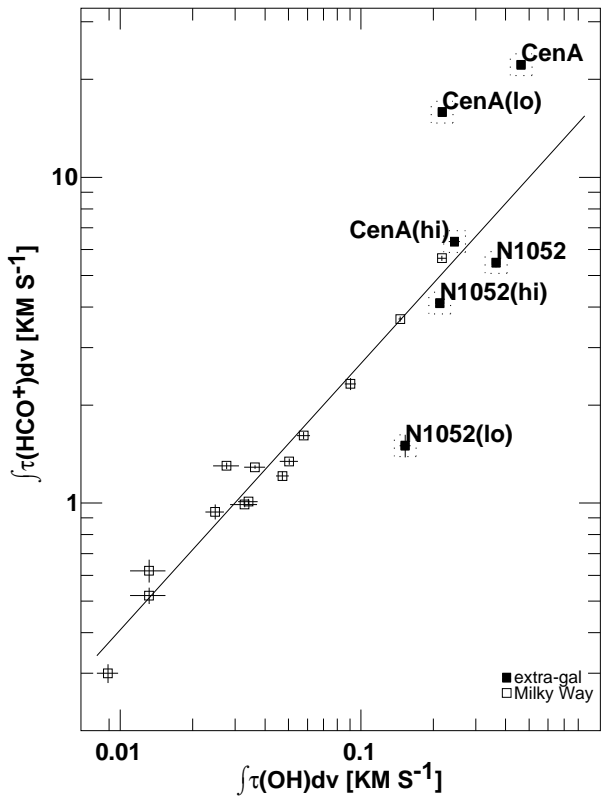,height=10.5cm}
\caption[]{OH and \hcop\ optical depths for galactic diffuse and 
translucent clouds, and elliptical galaxy lines of sight. The 
qualifications ``lo'' and ``hi'' in NGC 1052 and Cen A refer to 
the velocity ranges defined in Tables 1-2.  
}
\end{figure}

\subsection{Comparison with Cen A} 


Shown in Fig. 2 are profiles toward Cen A like those shown in Fig. 1 
for NGC 1052.  As toward NGC 1052, the absorption is red-shifted with 
respect to the systemic velocity, the most extreme velocities (relative 
to the systemic velocity) are more prominent in molecular gas, and 
several very prominent, discrete H I features have no counter part in 
molecular absorption.   The somewhat eccentric placement of the 
various narrow OH and \hcop\ components with respect to H I are
entirely typical of local lines of sight through the Milky Way 
\citep{LucLis96,LisLuc96}.

Until
recently it appeared that H I absorption was considerably less 
extended in velocity, but the broad, faint wing above 600 \kms\
discovered by  \citep{SarTro+02} corrects this misimpression.
Despite its substantial width, all of the absorption toward Cen A 
is much narrower than even the redshifted portion of the CO emission 
profile, which represents contributions from both the dust lane 
and inner circumnuclear disk \citep{Lis01a}.

\subsection{OH and \hcop\ in NGC1052, Cen A and the Milky Way}

Figure 3 shows the \hcop\ and OH optical depths which have
been observed in NGC 1052 and Cen A, along with the sample
of local diffuse clouds of \cite{LisLuc96} and \cite{LucLis00}.
As is clearly the case in Fig. 3, OH and \hcop\ absorption are 
extremely well correlated at low-to-moderate column densities.  
For reference, note that the OH column density toward \zoph, 
N(OH) $= 5 \times10^{13}~\pcc$, corresponds to an integrated 
opacity of approximately 0.05 \kms.  Thus the low-velocity 
profile integral toward NGC 1052 corresponds to little more than
the amount of molecular material in a single local diffuse cloud. 

The correlation between OH and \hcop\ column densities implied
by our data was shown by \cite{KanChe02} to extend in rather 
spectactular fashion to four sources observed at redshifts up 
to about unity.  By contrast, the OH/\hcop\ integrated optical 
depth ratios seen in NGC 1052 and Cen A vary substantially from 
source to source and across
the profile of either source individually.  The profile integrals
for lower and higher velocity gas (as defined in the footnotes
to Tables 1 and 2) are shown in Fig. 3 along with the ratio
of total integrals.  In this plot we see that both galaxies have
very similar amounts of OH absorption, but their \hcop\ integrals
differ by a factor of 4.  The \hcop/OH ratios vary by similar 
factors of 4-5 across the profile in both systems, but the overall 
ratio in NGC 1052 is close to that seen in diffuse gas locally 
while that in Cen A is much larger.  Not plotted, but as given in
Tables 1-2, the OH/H I optical depth ratios actually vary little
across the profile in either source, yet the OH/H I ratio
overall varies by a factor of 6 between the two systems, even
more than for OH/\hcop.  For nearly the same amount of OH absorption,
NGC 1052 has much smaller H I and \hcop\ profile integrals. 

For comparison we further note that the galactic OH absorption seen toward
Sgr A$^\ast$ in the VLA synthesis of \cite{KilLo+92} 
(at v $>$ -70 \kms, which omits some weak absorption arising in 
the -135 \kms\ expanding molecular ring feature) 
has an integrated optical depth 14.6 \kms.  This is 30-40 times higher 
than toward either elliptical.   However the Milky Way CO emission 
toward Sgr A$^\ast$ has a profile integral of 1000 K in the J=3-2 line 
\citep{LisBur95}, despite being heavily absorbed by intervening
gas (the J=1-0 profile integral from the 12m telescope is 1400 K \kms). 
Thus the CO emission/OH absorption ratio in the Milky
Way, 1000 K \kms/14 \kms $\simeq 70$ K across the entire profile, 
is actually quite similar to that in Cen A, 66 K \kms/0.463 \kms 
= 143 K.
 
\section{Summary}

The Plateau de Bure's detection of mm-wave absorption from \hcop, 
HCN and CO toward the center of the elliptical galaxy, LINER- and 
\HH O megamaser-host NGC 1052  provides new opportunities to probe 
the material 
which obscures and reddens the optical nucleus. OH and \hcop\ 
yield consistent estimates of the hydrogen column density 
in molecular gas 2N(\HH) $\simeq 5\times10^{21}~\pcc$, assuming
conditions like those seen locally in the Milky Way.  These
estimates are quite consistent with the amount of reddening
observed toward the nucleus (0.42 mag).

Our \hcop\ profile is somewhat broader than either H I or OH. Although 
chemical differences between OH and \hcop\ are not expected, a wide 
range of other influences may be called up on account for differences
in linewidths and line ratio variations across the profile.  These
include  excitation effects, secular evolution of the background
between epoochs of observation, and frequency-dependence of both the 
jet structure and the free-free opacity obscuring the background 
radio source.

CO emission has been unsuccessfuly sought (by others) toward 
NGC1052 at fairly low levels.  The absence of CO emission from 
gas having an optical depth of 0.1, as observed, and a molecular
column density as inferred, indicates that the molecular gas
distribution is small with respect to the singledish radiotelescope 
beams which have been employed. Current limits on CO emission
brightness correspond to received flux levels which are not
large compared to the amount of continuum flux which is being
absorbed.  As for H I, attempts to observe CO emission toward the
nucleus of NGC 1052 are seriously compromised by contamination from 
absorption of comparable strength.

The OH/\hcop\ ratio in NGC1052 overall is quite close to the value 
established for local diffuse clouds; the gas in NGC1052 is likely 
diffuse with a large fraction of molecular hydrogen but very incomplete
conversion of carbon into CO.  The Galactic OH/\hcop\ ratio has
also been found in several sources at redshift up to about 1.  Yet, 
the OH/\hcop\ ratio is markedly larger over the lower-velocity portion 
of the NGC1052 absorption line profile.  A comparison with 
observations of the other elliptical showing a similar wealth of 
molecular absorption lines, Cen A (NGC\\,5128) , shows that the OH/\hcop\ 
ratio also varies widely across the Cen A profile.  Overall, Cen A 
has much more H I and \hcop\ absorption than does NGC 1052, for
very nearly the same OH absorption. The OH profile integral 
observed in the Milky Way toward 
Sgr A$^\ast$ is 30-40 times larger than that observed toward either 
NGC 1052 or NGC 5128.  The reddening observed toward NGC1052 is
smaller by a similar factor (40) than that inferred toward Sgr A$^\ast$ .

Clarification of the distribution and kinematics of the molecular
gas in the nucleus would be aided by VLB maps at frequencies
comparable to (or above) those of the mm-wave absorption lines
we have detected, and by searches for CO emission in regions 
of the dust lane seen away from the nuclear radiocontinuum.

\begin{acknowledgements}

The National Radio Astronomy Observatory is operated by AUI, Inc. under 
a cooperative agreement with the US National Science Foundation.  IRAM 
is operated by CNRS (France), the MPG (Germany) and the IGN (Spain).  We 
owe the staff from IRAM and at the Plateau de Bure our thanks for their 
assistance in taking the data.  We thank Francoise Combes, Huub van 
Langevelde, Michael Rupen, Neil Killeen and Rene Vermeulen for 
providing their data in machine readable form.  Comments by the 
referee helped us to clarify the discussion of several important 
issues.

\end{acknowledgements}


\end{document}